\documentstyle[twoside,fleqn,espcrc2]{article}

\begin{document}
\hyphenation{Super-Kamiokande}
\input psfig

\title{SOLAR NEUTRINOS:WHAT NEXT?}

\author{J. N. BAHCALL\address{Institute for Advanced Study,
        Princeton, NJ 08540, USA\\E-mail: jnb@sns.ias.edu}}

\begin{abstract}
I summarize the current state of solar neutrino research and then
give my answer to the question: What should we do next?
\end{abstract}

\maketitle

\section{Introduction}

The reader who is familiar with solar neutrino research may
wish to skip directly to the last section entitled: What Next?

Solar neutrinos have been detected experimentally with fluxes and
energies that are qualitatively consistent with solar models that are
constructed assuming that the sun shines by nuclear fusion reactions.
The first experimental result, 
obtained by Ray Davis and his collaborators in
1968,~\cite{davis68,chlorine} 
has now been confirmed by four other beautiful experiments,
Kamiokande,~\cite{kamiokande} SAGE,~\cite{sage} GALLEX,~\cite{gallex}
and SuperKamiokande.~\cite{superk}  
The observation of solar neutrinos with approximately the predicted
energies and fluxes establishes empirically the theory~\cite{bethe39} 
that main
sequence stars
derive their energy from nuclear fusion reactions in their interiors
and has inaugurated what we all hope will be a flourishing field of
observational neutrino astronomy. 

Although the calculated neutrino fluxes depend upon high powers of the
central temperature of the solar model, 
the experiments and the solar model theory are so precise that 
persistent quantitative discrepancies have existed between the model
predictions and the solar model calculations for over thirty years~\cite{bahcall68,bahcall89,BP98}

Important experiments are underway that will provide diagnostic
information about the physical properties of neutrinos that are
created in the center of the sun and detected on earth in really long
baseline experiments.  At this workshop, we will hear discussions of
the SuperKamiokande, SNO, BOREXINO, \hbox{HELLAZ}, HERON, ICARUS, LENS, and
KamLAND experiments.

I will discuss predictions of the combined standard model in the main
part of this review.  By `combined' standard model, I mean the
predictions of the standard solar model and the predictions of the
standard electroweak model.  We need a solar model to tell us how many
neutrinos of what energy are produced in the sun and we need an
electroweak theory to tell us how the number and flavor content of the
neutrinos are changed as they make their way from the center of the
sun to detectors on earth.  For all practical purposes, the standard
electroweak model states that nothing happens to solar neutrinos
after they are created in the deep interior of the sun.  Using
standard electroweak theory and fluxes from the standard solar model,
one can calculate the rates of neutrino interactions in different
terrestrial detectors with a variety of energy sensitivities.  The
combined standard model also predicts that the energy spectrum from a
given neutrino source should be the same for neutrinos produced in
terrestrial laboratories and in the sun and that there should not be
measurable time-dependences (other than the seasonal dependence caused
by the earth's orbit around the sun).  The spectral and temporal
departures from standard model expectations are expected to be small
in all currently operating experiments~\cite{bks98} and have not yet
yielded definitive results. Therefore, I will concentrate here on
inferences that can be drawn by comparing the total rates observed in
solar neutrino experiments with the combined standard model
predictions.

I will begin by reviewing in Section~\ref{best} the quantitative 
 predictions of the
combined standard solar model and then
describe in Section~\ref{threeproblems} 
the three solar neutrino problems that are 
established by the chlorine, Kamiokande, SAGE, GALLEX, and
SuperKamiokande experiments.  In Section~\ref{uncertainties}, I detail the
 uncertainties in the standard model predictions and then show in
 Section~\ref{helioseismology} 
that helioseismological measurements indicate that the
 standard solar model predictions are accurate for our
 purposes. 
In Section~\ref{helioseismology}, 
I discuss the implications for solar neutrino
 research of the precise agreement between helioseismological
 measurements and the predictions of standard solar
 models. Next, ignoring all knowledge of the sun, I cite analyses  in
 Section~\ref{nomodels} that show that one cannot fit 
 the existing experimental data with neutrino fluxes that are  arbitrary
 parameters, unless one invokes new physics to change the shape or
 flavor content of the neutrino energy spectrum.
I summarize in Section~\ref{oscillations} the characteristics of the
 best-fitting neutrino oscillation descriptions of 
 the experimental data.
Finally, I will discuss and summarize the results in
 Section~\ref{discussion}. 

If you want to obtain  numerical data or subroutines 
that are discussed in this
talk, or to  see relevant background information, 
you can copy  them from my Web site: http://www.sns.ias.edu/$\sim$jnb.

\begin{figure}[!t]
\centerline{\psfig{figure=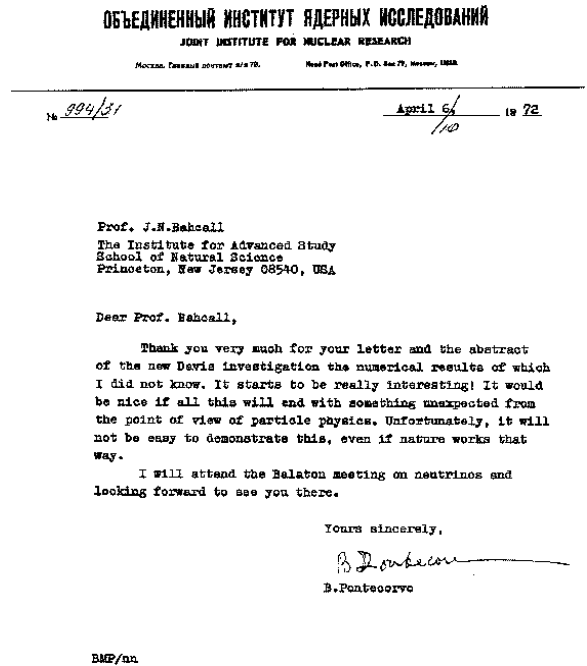,width=3in}}
\vglue-.3in
\caption[]{Letter from Bruno Pontecorvo in 1972.\label{fig:pontecorvo}}
\end{figure}

Before we begin the detailed discussion, I want to make just a brief
historical diversion.  Nearly all of the current interest in solar
neutrinos centers around the opportunity to use the sun as a neutrino
source in a very long baseline oscillation experiment.  In preparing
for a talk in honor of Fred Reines a few months ago, I ran across a
long forgotten 1972 letter from Bruno Pontecorvo, the originator of the
hypothesis that oscillations may be observed in solar neutrino
experiments. For your interest, I enclose a reproduction of this
letter in Fig.~\ref{fig:pontecorvo}.

For the benefit of 
neutrino pioneers of today, it is perhaps worth remarking that
Ray Davis and I never considered the possibility that solar neutrinos
could be used to learn more about neutrinos when, in the early 1960's, we 
were first analyzing the potentialities of a practical
chlorine experiment.  We sold the experiment as a fundamental  test of
 the hypothesis that the sun shines by nuclear fusion
reactions in its interior.  Only after the first results of the
chlorine experiment showed in 1968 a conflict with the solar model
calculations and Gribov and Pontecorvo published their epochal 1969 paper
on vacuum oscillations of solar neutrinos did we begin to consider the
possibility that solar neutrinos might tell us something new about
particle physics. Maybe there are previously unimagined physics
treasures to be discovered in future neutrino experiments.

\section{Standard Model Predictions}
\label{best}

\begin{table}[!t]
\centering
\begin{minipage}{3in}
\renewcommand{\arraystretch}{1.3}
\caption[]{Standard Model Predictions (BP98): 
solar neutrino fluxes and neutrino capture rates, with $1\sigma$
uncertainties from all sources (combined quadratically).
\protect\label{tab:bestestimate}}
\begin{tabular}{@{\extracolsep{-7pt}}|llcc|}
\hline
Source&\multicolumn{1}{c}{Flux}&Cl&Ga\\
&\multicolumn{1}{c}{$\left(10^{10}\ {\rm cm^{-2}s^{-1}}\right)$}&(SNU)&(SNU)\\[5pt]
\hline
pp&$5.94 \left(1.00^{+0.01}_{-0.01}\right)$&0.0&69.6\\
pep&$1.39 \times 10^{-2}\left(1.00^{+0.01}_{-0.01}\right)$&0.2&2.8\\
hep&$2.10 \times 10^{-7}$&0.0&0.0\\
${\rm ^7Be}$&$4.80 \times 10^{-1}\left(1.00^{+0.09}_{-0.09}\right)$&1.15&34.4\\
${\rm ^8B}$&$5.15 \times 10^{-4}\left(1.00^{+0.19}_{-0.14}\right)$&5.9&12.4\\
${\rm ^{13}N}$&$6.05 \times
10^{-2}\left(1.00^{+0.19}_{-0.13}\right)$&0.1&3.7\\
${\rm ^{15}O}$&$5.32 \times
10^{-2}\left(1.00^{+0.22}_{-0.15}\right)$&0.4&6.0\\
${\rm ^{17}F}$&$6.33 \times
10^{-4}\left(1.00^{+0.12}_{-0.11}\right)$&0.0&0.1\\[7pt]
Total&&$7.7^{+1.2}_{-1.0}$&$129^{+8}_{-6}$\\
\hline
\end{tabular}
\end{minipage}
\end{table}

\begin{figure}[!ht]
\hglue-.27in\psfig{figure=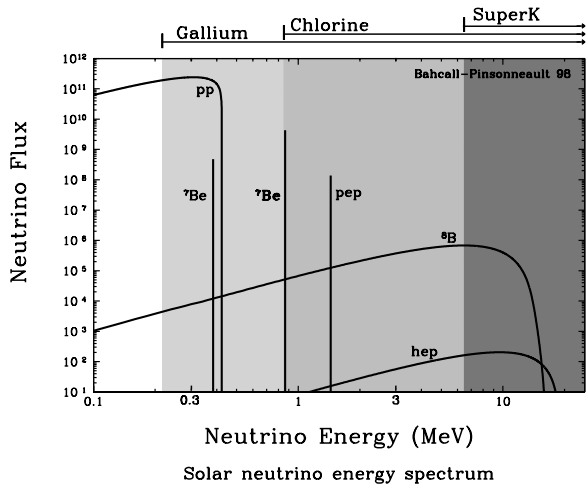,width=3.6in,angle=270}
\vglue-.3in
\caption[]{The energy Spectrum of neutrinos from the pp chain of 
interactions in the Sun, as predicted by the standard solar model.
Neutrino fluxes from continuum sources (such as pp and 8B) are given in the units
of counts per cm2 per second. The pp chain is responsible for more than 98\%
of the energy generation in the standard solar model. Neutrinos produced in the
carbon-nitrogen-oxygen CNO chain are not important energetically and are
difficult to detect experimentally. The arrows at the top of the figure indicate
the energy thresholds for the ongoing neutrino experiments.}
\label{fig:energyspectrum}
\end{figure}

Table~\ref{tab:bestestimate} gives the  neutrino fluxes and their 
uncertainties for our best standard solar model,  hereafter BP98.~\cite{BP98}
Figure~\ref{fig:energyspectrum} shows the predicted neutrino fluxes
from the dominant $p$-$p$ fusion chain.

The BP98  solar model includes diffusion of heavy elements and helium, 
makes use of the nuclear reaction rates recommended by the expert
workshop held at the Institute of Nuclear Theory,~\cite{adelberger98} 
recent (1996) Livermore
OPAL radiative opacities,~\cite{opacity} 
the OPAL equation of state,~\cite{eos}   and 
electron and ion screening as determined by the 
recent density matrix calculation.~\cite{gruzinov,salpeter}
The neutrino absorption cross sections that are used in constructing
Table~\ref{tab:bestestimate} are  the most
accurate  values available~\cite{bahcall97,boronspectrum} and include,  where appropriate,
the thermal energy of  fusing
solar ions  and 
improved nuclear and atomic data.
The validity of the absorption cross sections has recently been
confirmed experimentally using intense radioactive sources of ${\rm ^{51}Cr}$.
The ratio, $R$, of the capture rate 
measured (in GALLEX and SAGE) to the calculated ${\rm ^{51}Cr}$ capture rate
is $R = 0.95 \pm 0.07~{\rm (exp)}~+~^{+0.04}_{-0.03}~{\rm (theory)}$
and was discussed extensively at Neutrino 98 by Gavrin and by Kirsten.
The neutrino-electron scattering cross sections, used  in
interpreting the Kamiokande and SuperKamiokande experiments,
now include electroweak radiative
corrections.~\cite{sirlin} 

\begin{figure}[!ht]
\centerline{\psfig{figure=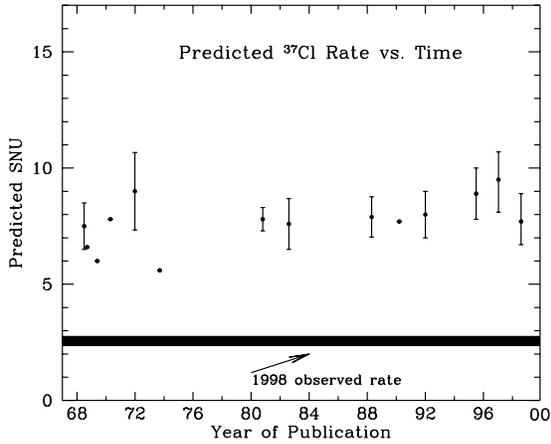,width=3.5in,angle=270}}
\vglue-.5in
\caption[]{The predictions of John Bahcall and his collaborators of
neutrino capture rates in the $^{37}$Cl experiment are
shown as a function of the date of publication (since the first
experimental report in 1968.~\cite{davis68})
The event rate SNU is
a convenient product of neutrino flux times interaction cross section,
$10^{-36}~{\rm interactions~per~target~atom~per~sec}$.
The format is from Figure 1.2 of the book Neutrino
Astrophysics.~\cite{bahcall89} The predictions have been updated through
1998.}
\label{fig:chlorine} 
\end{figure}

Figure~\ref{fig:chlorine} shows for the chlorine experiment 
all the predicted rates and the
estimated uncertainties ($1\sigma$) published by my colleagues and myself since
the first measurement by Ray Davis and his colleagues in 1968.
This figure should give you some feeling for the robustness of the
solar model calculations. Many
hundreds and probably thousands of researchers 
have, over three
decades, made great improvements in  the input data for the solar
models, including nuclear cross sections, neutrino cross sections, 
measured element
abundances on the surface of the sun, the solar luminosity, the stellar
radiative 
opacity, and the stellar equation of state. Nevertheless, the most accurate
predictions of today are essentially the same as they were in 1968
 (although now they can be made with much greater confidence).
For the gallium experiments, the 
neutrino fluxes predicted by standard solar models, 
corrected for
diffusion,  have been in the 
range $120$ SNU to $141$ SNU since 1968.~\cite{bahcall97}
A SNU is a convenient unit with which to describe the
measured rates of solar neutrino experiments: $10^{-36}$ interactions
per target atom per second.

There are three reasons that the theoretical calculations of 
neutrino fluxes are robust: 1) the availability of precision 
measurements and precision 
calculations of input data; 2)
the connection between neutrino fluxes and the measured solar
luminosity; and 3) the measurement of the helioseismological  frequencies
of the solar pressure-mode ($p$-mode) eigenfrequencies.
I have discussed these reasons in detail in another talk.~\cite{heineman}

\begin{figure}[!b] 
\centerline{\psfig{figure=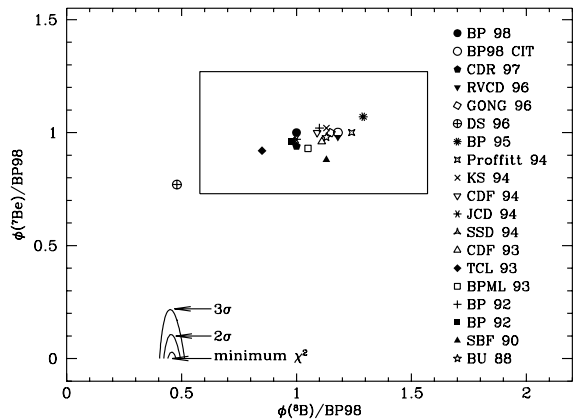,width=3in}}
\vglue-.3in
\caption[]{Predictions of standard solar models since 1988.
This figure, which is Fig.~1 of Bahcall, Krastev and 
Smirnov (1998)\cite{bks98},
shows the predictions of $19$ standard solar models in the
plane defined by the ${\rm ^7Be}$ and ${\rm ^8B}$ neutrino fluxes. 
The abbreviations that are used in the figure to identify different
solar models are defined in the bibliographical item,
Ref.~\cite{models}.
The figure includes all standard solar models with which I am  familiar that
were published in refereed journals
in the decade 1988-1998.
All of the
fluxes are normalized to the predictions of the 
Bahcall-Pinsonneault 1998 solar model, BP98.~\cite{BP98}
The rectangular error box defines the $3\sigma$ error range of the
BP98 fluxes. The best-fit ${\rm ^7Be}$ neutrino flux is negative. At the
$99$\% C.L., there is no solution~\cite{bks98} 
with all positive neutrino fluxes (see discussion in Section~\ref{nomodels}).
All of the standard model solutions lie  far from the best-fit
solution, even far from the $3\sigma$ contour.}
\label{fig:independent}
\end{figure}

Figure~\ref{fig:independent}
displays the 
calculated  ${\rm ^7Be}$ and ${\rm
^8B}$ neutrino fluxes 
for all $19$ standard solar models 
which have been published in the last $10$ years in refereed
science journals.  
The fluxes are normalized by 
dividing each
published value by the flux from the BP98
solar model;~\cite{BP98} the abscissa is the
normalized ${\rm ^8B}$ flux and the ordinate
 is the normalized ${\rm ^7Be}$
neutrino flux.  The rectangular box shows the estimated 
$3\sigma$ uncertainties in the
predictions of the BP98 solar model.

All of the solar model results from different groups 
fall within the estimated 3$\sigma$
uncertainties in the BP98 analysis (with the exception of the
Dar-Shaviv model whose results have not been reproduced by other groups).
This agreement 
demonstrates the robustness of the predictions since the
calculations use different computer codes (which achieve varying
degrees of precision) and 
involve a variety of choices for the nuclear
parameters, the equation of state, the stellar radiative opacity, 
the initial heavy element abundances, and the physical processes
that are included.

The largest contributions to  the dispersion in values in
Figure~\ref{fig:independent} are due to the
choice of the normalization for  
$S_{17}$ (the production
cross-section factor for ${\rm ^8B}$ neutrinos) and the
inclusion, or non-inclusion, of element diffusion in the 
stellar evolution codes.
The effect in the plane of Fig.~\ref{fig:independent} 
of the normalization of
$S_{17}$ is shown by the difference between the point for BP98
 (1.0,1.0), which was computed using the most recent recommended 
 normalization,~\cite{adelberger98} and the
point at (1.18,1.0) which corresponds to the BP98 result with the
earlier (CalTech) normalization.~\cite{CIT}   

Helioseismological-observations have shown~\cite{BP98,prl97} that
element diffusion is occurring and must be included in solar models,
so that the most recent models shown in Fig.~\ref{fig:independent} now
all include helium and heavy element diffusion.  By comparing a large
number of earlier models, it was shown that all published standard
solar models give the same results for solar neutrino fluxes to an
accuracy of better than 10\% if the same input parameters and physical
processes are included.~\cite{BP92,BP95}

Bahcall, Krastev, and Smirnov~\cite{bks98} have compared the observed rates 
with the calculated, standard model values, combining  quadratically
the theoretical solar model and experimental
uncertainties, as well as the uncertainties in the neutrino cross
sections.
Since the GALLEX and SAGE experiments measure the same quantity, we
treat the weighted average rate in gallium as one experimental number.
We adopt the SuperKamiokande measurement as the most precise direct
determination of the higher-energy ${\rm ^8B}$ neutrino flux.

\begin{figure}[!t]
\hglue-.48in\psfig{figure=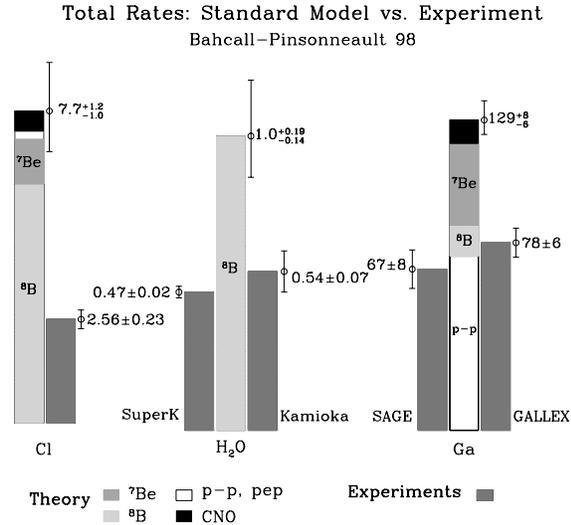,width=4.1in,angle=270}
\vglue-.5in
\caption[]{Comparison of measured 
rates and 
standard-model predictions
for five solar neutrino 
experiments.~\cite{chlorine,kamiokande,sage,gallex,superk}
The unit for the radiochemical experiments (chlorine and gallium) is
SNU (see Fig.~\ref{fig:chlorine} for a definition); the unit for the
water-Cerenkov experiments (Kamiokande and SuperKamiokande) is the rate
predicted by the standard solar model plus standard electroweak
theory.~\cite{BP98} 
\label{fig:compare}}
\end{figure}

Using the predicted fluxes from the BP98
model, the $\chi^2$ for the fit to the three  experimental rates
 (chlorine, gallium, and SuperKamiokande, see Fig.~\ref{fig:compare}) is

\begin{equation}
\chi^2_{\rm SSM} \hbox{(3 experimental rates)} = 61\ .
\label{chitofit}
\end{equation}
The result given in Eq.~(\ref{chitofit}), which is approximately 
equivalent to  a
$20\sigma$ discrepancy,  is a quantitative expression 
of the fact that the standard model predictions do not fit the
observed solar neutrino measurements.

\section{Three Solar Neutrino Problems}
\label{threeproblems}

I will  now compare  the predictions of the combined
standard model with the results of the  operating solar neutrino
experiments.    

We will see that this comparison leads to three
different discrepancies between the calculations and the observations,
which I will refer to as the three solar neutrino problems.

Figure~\ref{fig:compare} shows  the measured and the calculated event
rates in the five ongoing solar neutrino experiments.  This figure
reveals three discrepancies between the 
experimental results and the expectations based
upon the combined standard model.  As we shall see, only
the first of these discrepancies depends in an important way upon the 
predictions of the standard solar model.

\subsection{Calculated versus Observed Absolute Rate}
\label{firstproblem}

The first solar neutrino experiment to be performed was the chlorine
radiochemical experiment,~\cite{chlorine} 
which detects electron-type neutrinos that
are  more energetic 
than $0.81$ MeV.  After more than
a quarter of a century of operation of this experiment, the measured event
rate is $2.56 \pm 0.23$ SNU, which is a factor of three less than
is predicted by the most detailed theoretical calculations,
$7.7_{-1.0}^{+1.2}$ SNU.~\cite{BP98}  
Most of the predicted rate in the chlorine experiment is
from the rare, high-energy $^8$B neutrinos, although the $^7$Be
neutrinos are also expected to contribute significantly.  According to
standard model calculations, the $pep$ neutrinos and the CNO neutrinos 
 (for simplicity not discussed here)
are expected to contribute less than 1 SNU to the
total event rate.

This discrepancy between the calculations and the observations for the
chlorine experiment was, for more than two decades, the only solar
neutrino problem. I shall refer to the chlorine disagreement 
as the ``first'' solar neutrino
problem.

\subsection{Incompatibility of Chlorine and Water 
Experiments}

The second solar neutrino problem results from a comparison of the 
measured event rates in the chlorine experiment and in the Japanese
pure-water experiments,  Kamiokande~\cite{kamiokande} and 
SuperKamiokande.~\cite{superk}  
The water experiments detect
higher-energy neutrinos, most easily  above $7$ MeV,
by by observing the Cerenkov radiation from 
neutrino-electron scattering: $\nu ~+~e ~\longrightarrow
  \nu' ~+~e'.$   According to the standard solar model, 
\hbox{$^{8}$B} beta decay, and possibly the $hep$ reaction,~\cite{bk98} 
are the only important 
source of these higher-energy neutrinos. 

The Kamiokande and SuperKamiokande experiments show
 that the observed neutrinos come from
the sun. 
The  electrons that are scattered by the incoming neutrinos recoil
predominantly  in the direction of the sun-earth vector;  the
relativistic electrons are observed by the 
Cerenkov radiation they produce in the water detector.
In addition, the water Cerenkov 
experiments measure
the energies of individual scattered electrons and therefore
provide information about the energy spectrum of the incident 
solar neutrinos.

The total event rate in the water  experiments, about $0.5$ the 
standard model value (see Fig.~\ref{fig:compare}), is 
determined by the same high-energy $^8$B neutrinos that are expected,
on the basis of the combined standard model,
to dominate the event rate in the chlorine experiment.
I have  shown elsewhere~\cite{Bahcall91} that solar physics changes   
the shape of the \hbox{$^{8}$B} neutrino spectrum by less than  1 part
in $10^5$~. 
Therefore, we can calculate the rate in the chlorine experiment 
 (threshold $0.8$ MeV)
that
is produced by  
the \hbox{$^{8}$B} neutrinos observed
in the Kamiokande and SuperKamiokande experiments at an order of
magnitude higher energy threshold.

If no new physics changes the shape of the $^8$B neutrino energy spectrum,
the    chlorine rate from \hbox{$^{8}$B} alone
is  $2.8 \pm 0.1$ SNU for the SuperKamiokande normalization 
 ($3.2 \pm 0.4$ SNU for the 
Kamiokande normalization), which exceeds the total observed chlorine 
rate of $2.56 \pm 0.23$ SNU. 

Comparing the rates of the
SuperKamiokande and the chlorine experiments, one finds--assuming that
the shape of the energy  spectrum of  $^8$B  $\nu_e$'s is not changed
by new physics--that the net
contribution to the chlorine experiment from the $pep$, $^{7}$Be, and CNO
neutrino sources is negative: $-0.2 \pm 0.3$ SNU.
The contributions from the $pep$, $^7$Be, and CNO neutrinos 
would appear to be completely missing;  the standard model prediction for 
the combined contribution of $pep$, $^7$Be, and CNO neutrinos is a 
relatively large $1.8$~SNU (see Table~\ref{tab:bestestimate}).  
On the other hand, we know
that the $^7$Be neutrinos must be created in the sun since they are
produced by electron capture on the same isotope ($^7$Be) which
gives rise to the $^8$B neutrinos by proton capture. 

Hans Bethe and I pointed out~\cite{bethe90} 
that this apparent incompatibility of the
chlorine and water-Cerenkov experiments constitutes  a ``second'' 
solar neutrino problem that is almost independent of the absolute
rates predicted by solar models. 
The inference that is usually made from this comparison is that the
energy spectrum of ${\rm ^8B}$ neutrinos is changed from the standard
shape by physics not included in the simplest version of the standard
electroweak model.

\subsection{Gallium Experiments: No Room for $^{7}$Be Neutrinos}
\label{galliumproblem}

The results of the 
gallium experiments, GALLEX and SAGE,
constitute the third solar neutrino problem.
The average observed  rate in these two experiments is $73 \pm 5$ 
SNU, which
is accounted for in the standard model by the 
theoretical  rate of $72.4$ SNU
that is calculated to come from the basic $p$-$p$ and $pep$ neutrinos
 (with only a 1\% uncertainty in the standard solar model $p$-$p$ flux).
The \hbox{$^{8}$B} neutrinos, which are observed above $6.5$ MeV 
in the Kamiokande experiment, must also contribute to the gallium
event rate. 
Using the standard shape for the spectrum of ${\rm ^8B}$
neutrinos and normalizing to the rate observed in Kamiokande, 
${\rm ^8B}$ contributes  another $6$ SNU. (The contribution predicted by the
standard model is
$12$~SNU, see Table~\ref{tab:bestestimate}.)  
Given the measured rates in the
gallium experiments, there is no room for the additional $34 \pm 3$
SNU that
is expected from $^{7}$Be neutrinos on the basis of 
standard solar models (see Table~\ref{tab:bestestimate}).

The seeming exclusion of everything but $p$-$p$ neutrinos in the gallium
experiments is the ``third'' solar neutrino problem.  This problem is
essentially independent of the  previously-discussed solar
neutrino problems, since it depends strongly 
upon the $p$-$p$ neutrinos that are not observed in
the other experiments and whose theoretical flux can be calculated
accurately.

The missing $^7$Be neutrinos cannot be
explained away by a change in solar physics. The \hbox{$^{8}$B}
neutrinos that are observed in the Kamiokande experiment are produced
in competition with the missing $^7$Be neutrinos; 
the competition is between electron capture on $^7$Be versus
proton capture on $^7$Be.
Solar model
explanations that reduce the predicted ${\rm ^7Be}$ flux 
generically reduce much
more (too much) the predictions for the observed ${\rm ^8B}$ flux.

\begin{table*}[t!]
\centering
\renewcommand{\arraystretch}{1.3}
\caption[]{Average uncertainties in neutrino fluxes and event rates 
due to different input data.  The flux uncertainties are expressed in
fractions of the total flux and the event rate uncertainties are
expressed in SNU.  The ${\rm ^7Be}$ electron capture rate causes an
uncertainty of $\pm 2\%$~\cite{be7paper} that affects only the ${\rm
^7Be}$ neutrino flux.  The average fractional uncertainties for
individual parameters are shown.
\protect\label{tab:uncertainties}}
\begin{tabular}{|l@{\extracolsep{-5pt}}c@{\extracolsep{-4pt}}cccccccc|}
\hline
$<$Fractional&pp&${\rm ^3He ^3He}$&${\rm ^3He ^4He}$&${\rm ^7Be} +
p$&$Z/X$&opac&lum&age&diffuse\\
uncertainty$>$&\ \ \ 0.017\ \ \ \ \ &0.060&0.094&0.106&0.033&{see text}&0.004&0.004&0.15\\
\hline
Flux&&&&&&&&&\\ \cline{1-1}
pp&0.002&0.002&0.005&0.000&0.002&0.003&0.003&0.0&0.003\\
${\rm ^7Be}$&0.0155&0.023&0.080&0.000&0.019&0.028&0.014&0.003&0.018\\
${\rm ^8B}$&0.040&0.021&0.075&0.105&0.042&0.052&0.028&0.006&0.040\\
SNUs&&&&&&&&&\\ \cline{1-1}
Cl&0.3&0.2&0.5&0.6&0.3&0.4&0.2&0.04&0.3\\
Ga&1.3&0.9&3.3&1.3&1.6&1.8&1.3&0.20&1.5\\
\hline
\end{tabular}
\end{table*}

The flux of $^7$Be neutrinos, $\phi({\rm ^7Be})$, is independent of
measurement uncertainties in  the
cross section for the nuclear reaction
${\rm ^7Be}(p,\gamma)^8$B; the  cross
section for this proton-capture  reaction is  the most uncertain
quantity that enters  in an important way in the solar
model calculations.  The flux of $^7$Be neutrinos depends upon the
proton-capture
reaction only through the ratio
\begin{equation}
\phi({\rm ^7Be}) ~\propto~ {{R(e)} \over {R(e) + R(p)}} ,
\label{Beratio}
\end{equation}
where $R(e)$ is the rate of electron capture by $^7$Be nuclei and
$R(p)$ is the rate of proton capture by $^7$Be.  With standard
parameters, solar models yield $R(p) \approx 10^{-3} R(e)$.
Therefore, one would have to increase the value of the ${\rm
^7Be}(p,\gamma)^8$B cross section by more than two orders of magnitude
over the current best-estimate (which has an estimated experimental
uncertainty of \hbox{$\sim$ 10\%}) in order to affect significantly
the calculated $^7$Be solar neutrino flux.  The required change in the
nuclear physics cross section would also increase the predicted
neutrino event rate by more than 100 in the Kamiokande experiment,
making that prediction completely inconsistent with what is observed.

I conclude that either: 1) 
at least three of the five operating solar neutrino
experiments (the two gallium experiments plus either chlorine or the
two water Cerenkov experiments,
Kamiokande and SuperKamiokande) 
 have yielded misleading results, or 2) physics beyond the standard
electroweak model is required to change the energy spectrum of $\nu_e$
after the neutrinos are produced in the center of the sun.

\section{Uncertainties in the Flux Calculations}
\label{uncertainties}

I will now  discuss  uncertainties in the
solar model flux calculations.

Table~\ref{tab:uncertainties} 
summarizes the uncertainties in the most important
solar neutrino fluxes and in the Cl and Ga event rates 
due to different nuclear fusion reactions (the
first four entries), the heavy element to hydrogen mass ratio (Z/X),
the radiative opacity, the solar luminosity, the assumed solar age,
and the helium and heavy element diffusion coefficients. 
The ${\rm ^{14}N} + p$ reaction causes a
0.2\% uncertainty in the predicted pp flux and a 0.1 SNU uncertainty
in the Cl (Ga) event rates.

The predicted event rates for
the chlorine and gallium experiments use recent improved calculations
of neutrino
absorption cross sections.~\cite{bahcall97,boronspectrum} 
The uncertainty in the
prediction for the gallium rate is dominated by  uncertainties in the
neutrino absorption cross sections, $+6.7$ SNU ($7$\% of the predicted
rate) and $-3.8$ SNU ($3$\% of the predicted rate). The
uncertainties in the chlorine absorption cross sections cause an
error, $\pm 0.2$ SNU ($3$\% of the predicted rate), that is relatively
small compared to other
uncertainties in predicting the rate 
for this experiment.  For non-standard neutrino energy
spectra that result from new neutrino physics, the uncertainties in
the predictions 
for currently favored solutions (which reduce the contributions from
the least well-determined $^8$B neutrinos) will
in general be less than the values quoted here for standard spectra
and must be calculated using the appropriate cross section uncertainty
for each neutrino energy.~\cite{bahcall97,boronspectrum}

The nuclear fusion
uncertainties in Table~\ref{tab:uncertainties} were taken  from
Adelberger et al.,~\cite{adelberger98} 
the neutrino 
cross section uncertainties from Bahcall (1997)\cite{bahcall97} 
and Bahcall et al. (1996),\cite{boronspectrum}
the heavy element uncertainty was taken from
helioseismological measurements,~\cite{basu97} the luminosity
and age uncertainties were adopted from BP95,~\cite{BP95}  the
$1\sigma$ fractional uncertainty in the diffusion rate was taken to be
$15$\%,~\cite{thoul} which is supported by helioseismological evidence,~\cite{prl97} and the opacity uncertainty was determined by comparing
the results of fluxes computed using the older Los Alamos opacities
with fluxes computed using the modern Livermore opacities.~\cite{BP92} 
 To include the effects of asymmetric errors, the now publicly-available
code for calculating rates and uncertainties (see discussion in
previous section)
was run with different input uncertainties and the results averaged.
The software contains a description of how each of the uncertainties
listed in Table~\ref{tab:uncertainties} were determined and used.

The low energy cross section
of the ${\rm ^7Be} + p$ reaction is the most important quantity that must be
determined more accurately in order to decrease the error in the
predicted event rates in solar neutrino experiments.
The $^8$B neutrino flux that is measured by the
Kamiokande,~\cite{kamiokande} 
Super-Kamiokande,~\cite{superk} and SNO~\cite{sno} 
experiments is, in all standard solar model calculations, 
directly proportional to the ${\rm ^7Be} + p$ cross section. 
If the $1\sigma$ uncertainty in this cross section can be reduced
by a factor of two to 5\%, then it will no longer be the limiting
uncertainty in predicting the crucial $^8$B neutrino flux
 (cf.~Table~\ref{tab:uncertainties}).

\section{How Large an Uncertainty Does Helioseismology Suggest?}
\label{helioseismology}

Could the solar model calculations be wrong by enough to explain the
discrepancies between predictions and measurements for solar neutrino
experiments?  Helioseismology, which confirms predictions of the
standard solar model to high precision, suggests that the answer is
probably ``No.''

\begin{figure}[!th]
\centerline{\psfig{figure=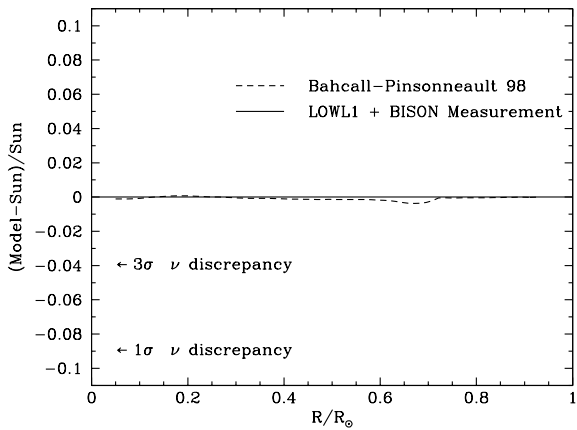,width=3.3in,angle=270}}
\vglue-.4in
\caption[]{Predicted versus Measured Sound Speeds. 
This figure shows
the excellent agreement between the calculated (solar model BP98, Model) 
 and the measured (Sun) sound
speeds, a fractional difference of $0.001$ rms for all 
speeds measured between $0.05 R_\odot$
and $0.95 R_\odot$. The vertical scale is chosen so as 
to emphasize that the fractional error is much smaller than generic
changes in the model, $0.04$ to $0.09$, that might
significantly affect the solar neutrino predictions.
\protect\label{modelsunnu}}
\end{figure}

Figure~\ref{modelsunnu} shows the fractional differences 
between the most accurate 
available sound speeds
measured by helioseismology~\cite{heliobest} and 
sound speeds calculated with our best
solar model (with no free parameters). 
The horizontal line corresponds to the hypothetical case in which the
model predictions exactly match the observed values.
The rms fractional 
difference between the
calculated and the measured sound speeds is $1.1\times10^{-3}$ for the
entire region over which the sound speeds are measured, $0.05R_\odot <
R < 0.95 R_\odot$.  In the solar core, $0.05R_\odot <
R < 0.25 R_\odot$ (in  which about $95$\% of the solar energy 
and neutrino flux is
produced in a standard model), the rms fractional 
difference between measured and
calculated sound speeds is $0.7\times10^{-3}$.

Helioseismological measurements also determine 
two other parameters that help characterize
 the outer part of the sun (far from
the inner region in which neutrinos are produced): the depth of the
solar convective zone (CZ), the region in the outer part of the sun
that is fully convective, and the present-day surface abundance by mass of
helium ($Y_{\rm surf}$).  The measured values, $R_{\rm CZ} =
(0.713 \pm 0.001) R_\odot$,~\cite{basu95} and $Y_{\rm surf} = 0.249 
\pm 0.003$,~\cite{basu97}
are in satisfactory agreement with the values predicted by the solar
model BP98, namely, $R_{\rm CZ} =
0.714 R_\odot$, and $Y_{\rm surf} = 0.243$.
However, we shall see below that precision measurements of the sound
speed near the transition between the radiative interior (in which energy
is transported by radiation)  and the
outer convective zone (in which energy is transported by convection) 
reveal small discrepancies between the model
predictions and the observations in this region.

If solar physics were responsible for the solar neutrino problems, how
large would one expect the discrepancies to be between solar 
model predictions and
helioseismological observations?
The characteristic size of the discrepancies 
 can be estimated  using the results of the neutrino experiments and  
scaling laws for
neutrino fluxes and sound speeds.

All recently published solar models predict essentially 
the same fluxes from the fundamental
pp and pep reactions (amounting to $72.4$ SNU in gallium
experiments, cf.~Table~1), which are closely related to the solar
luminosity.  
Comparing the measured gallium rates 
and the standard predicted rate for the gallium
experiments, the $^7$Be flux must be reduced  by a factor $N$
if the disagreement is not to exceed $n$ standard deviations, where 
$N$ and $n$ satisfy  $72.4 + (34.4)/N = 72.2 + n \sigma$. For a $1
\sigma $ ($3\sigma$) disagreement, $ N = 6.1 (2.05)$.  
Sound
speeds scale like the square root of the local temperature divided by
the mean molecular weight
and the $^7$Be neutrino flux scales approximately as the $10$th power
of the temperature.~\cite{ulmer} 
Assuming that the temperature changes are dominant,
agreement to within $1\sigma$ would
require fractional changes of order $0.09$ in sound speeds ($3\sigma$
could be reached with $0.04$ changes), if all model changes were in the
temperature\footnote{I have used in this calculation the GALLEX and
SAGE measured rates reported by Kirsten and Gavrin at Neutrino 98. The
experimental rates used in BP98 were not as precise and therefore
resulted in slightly less stringent constraints than those imposed
here. In BP98, we found that agreement to within $1\sigma$ with the
then available experimental numbers would
require fractional changes of order $0.08$ in sound speeds ($3\sigma$
could be reached with $0.03$ changes.)}.
This argument is conservative because it ignores 
 the $^8$B and CNO neutrinos which contribute to the observed
counting rate (cf.~Table~1) and which, if included, would require an
even larger reduction of the $^7$Be flux. 

I  have chosen the vertical scale in Fig.~\ref{modelsunnu}
 to be appropriate for
fractional differences between measured and predicted 
sound speeds that are of order $0.04$ to $0.09$ and 
that might therefore affect solar neutrino calculations. 
Fig.~\ref{modelsunnu}
 shows that the characteristic agreement between solar model
predictions and helioseismological measurements is more than a factor
of $40$ better than would be expected if there were a solar model
explanation of the solar neutrino problems.

\section{Fits Without Solar Models}
\label{nomodels}

Suppose 
(following the precepts of Hata et al.,~\cite{hata94} Parke,~\cite{parke95}
 and Heeger and Robertson~\cite{robertson})
we now ignore everything we  have learned  about  solar models over
the last $35$ years and allow
the important $pp$,
${\rm ^7Be}$, and ${\rm ^8B}$ fluxes to take on any non-negative
values.  What is the best fit that one can obtain
to the solar neutrino measurements assuming only
that the luminosity of the sun is supplied by nuclear fusion reactions among
light elements (the so-called `luminosity constraint')?~\cite{howwell}

The answer is that the fits are bad, even if we completely ignore what
we know about the sun. I quote the results from 
Bahcall, Krastev and Smirnov (1998).

If the CNO neutrino fluxes are set equal to zero,
there are no acceptable solutions at the $99$\% C. L. ($\sim 3\sigma$
result). The best-fit is worse if the CNO fluxes are not set
equal to zero.  All so-called `solutions' of the solar neutrino
problems in which the astrophysical model is changed arbitrarily
 (ignoring helioseismology and other constraints) are inconsistent with
the observations at much more than a $3\sigma$ level of significance.
No fiddling of the physical conditions in the model can yield the
minimum value, quoted above, that was 
 found by varying the fluxes independently and arbitrarily.

Figure~\ref{fig:independent} shows, in the lower left-hand corner,
 the best-fit solution and the
 $1\sigma $
 --$3\sigma $ contours.  The $1\sigma$ and $3\sigma$ limits  
were obtained by requiring
that $\chi^2 =  \chi^2_{\rm min} + \delta\chi^2 $, where for $1\sigma$
$\delta\chi^2 = 1 $ and for $3\sigma$ $\delta\chi^2 = 9$.  All of
 the standard 
model solutions lie far from the best-fit solution and  even lie far 
from the $3\sigma$ contour.

Since standard model descriptions do not fit the solar neutrino data, 
we will now consider
models  in which neutrino oscillations change the
shape of the neutrino energy spectra.

\begin{table}[t!]
\centering
\renewcommand{\arraystretch}{1.4}
\caption[]{Neutrino Oscillation Solutions.
\protect\label{tab:oscillatesolutions}}
\begin{tabular}{|llc|}
\hline
Solution&\multicolumn{1}{c}{$\Delta m^2$}&$\sin^2 2\theta$\\
\hline
SMA&$5 \times 10^{-6}~{\rm eV^2}$&$5 \times 10^{-3}$\\
LMA&$2 \times 10^{-5}~{\rm eV^2}$&0.8\\
LOW&$8 \times 10^{-8}~{\rm eV^2}$&0.96\\
VAC&$8 \times 10^{-11}~{\rm eV^2}$&0.7\\
\hline
\end{tabular}
\end{table}

\section{Neutrino Oscillations}
\label{oscillations}

The experimental results from all five of the operating solar neutrino
experiments (chlorine, Kamiokande, SAGE, GALLEX, and SuperKamiokande)
can be fit well by descriptions involving neutrino oscillations,
either vacuum oscillations (as originally suggested by Gribov and
Pontecorvo~\cite{vac})or resonant matter oscillations (as
originally discussed by Mikheyev, Smirnov, and Wolfenstein (MSW)~\cite{msw}).

Table~\ref{tab:oscillatesolutions} summarizes the four best-fit
solutions that are found in the two-neutrino
approximation.~\cite{bks98,bk98}  Only the SMA and vacuum 
oscillation solutions fit
well the  recoil electron energy spectrum measured
in the SuperKamiokande experiment--if the standard value for the $hep$
production reaction cross section (${\rm ^3He} ~+~ p 
~\rightarrow ~{\rm ^4He} ~+~e^+ ~+~\nu_e $) is used.~\cite{bks98} 
However, for over a decade
I have not given an estimated uncertainty for this cross
section.~\cite{bahcall89} 
The transition matrix element is essentially forbidden and the actual
quoted value for the production cross section depends upon a delicate
cancellation between two comparably sized terms that arise from very
different and hard to evaluate nuclear physics.  I do not see anyway
at present to determine from experiment or from first principles
theoretical calculations a relevant, robust upper limit to the $hep$
production cross section (and therefore the $hep$ solar neutrino flux).

The possible role of $hep$ neutrinos in solar neutrino experiments is
discussed extensively in Bahcall and Krastev (1998) ~\cite{bk98} 
The most important unsolved
problem in theoretical 
nuclear physics related to solar neutrinos is the range of
values allowed by fundamental physics for the $hep$ production cross
section.

\section{Discussion}
\label{discussion}

When the chlorine solar neutrino experiment was first
proposed,~\cite{davis64} the only stated motivation was ``...to see
into the interior of a star and thus verify directly the hypothesis of
nuclear energy generation in stars.''  This goal has now been
achieved, 

The focus has shifted to using solar neutrino experiments as a tool
for learning more about the fundamental characteristics of neutrinos
as particles.  Experimental effort is now concentrated on answering
the question: What are the probabilities for transforming a solar
$\nu_e$ of a definite energy into the other possible neutrino states?
Once this question is answered, we can calculate what happens to
$\nu_e$'s that are created in the interior of the sun.  Armed with
this information from weak interaction physics, we can return again to
the original motivation of using neutrinos to make detailed,
quantitative tests of nuclear fusion rates in the solar interior.
Measurements of the flavor content of the dominant low energy neutrino
sources, $p$-$p$ and $^7$Be neutrinos, will be crucial in this
endeavor and will require another generation of superb solar neutrino
experiments (see the comments in Section~\ref{whatnext}).

Three decades of refining the input data
and the solar model calculations has led to a predicted standard model
event rate for
the chlorine experiment, $7.7$ SNU, which is very close to $7.5$ SNU, the
best-estimate value 
obtained in 1968.~\cite{bahcall68} The
situation regarding solar neutrinos is, however, completely different
now, thirty years later.
Four experiments have confirmed the original chlorine 
detection of solar neutrinos.
Helioseismological measurements are in excellent agreement with the
standard solar model predictions and very strongly disfavor (by a
factor of $40$ or more) hypothetical deviations from the standard
model that are require to fits the neutrino data (cf.~Fig.~\ref{modelsunnu}).
Just in the last two years, 
improvements in the helioseismological measurements have 
 resulted in a five-fold improvement in the agreement between
the calculated standard solar model sound speeds and the measured
solar velocities (cf. Figure~2 of
the Neutrino~96 talk~\cite{neutrino96} with Figure~\ref{modelsunnu} of this
talk).

\section{What next?}
\label{whatnext}

More than $98$\% of the calculated standard model solar neutrino flux
lies below $1$ MeV. The rare $^8$B neutrino flux is the only solar
neutrino source for which measurements of the energy have been made,
but $^8$B neutrinos constitute a fraction of less than $10^{-4}$ of
the total solar neutrino flux.

The next goal of solar neutrino astronomy is to measure neutrino
fluxes below $1$ MeV. We should begin today preparing for experiments
that will measure the $^7$Be neutrinos (energy of $0.86$ MeV) and the
fundamental $p$-$p$ neutrinos ($< 0.43$ MeV).  Indeed, we have heard at
this workshop some marvelously exciting descriptions of how such low
energy experiments could be carried out.  The BOREXINO observatory,
which can detect $\nu-e$ scattering, is the only approved solar
neutrino experiment which can measure energies less than $1$ MeV.

The $p$-$p$  neutrinos are overwhelmingly the most abundant source of
solar neutrinos, carrying about $91$\% of the total flux according to
the standard solar model. The $^7$Be neutrinos constitute about $7$\%
of the total standard model flux.

If we want to test and to understand neutrino oscillations with high
precision using solar neutrino sources, then we have to measure the
neutrino-electron scattering rate with $^7$Be neutrinos, as will be
done with the BOREXINO experiment, and also the CC
(neutrino-absorption) rate with $^7$Be neutrinos (no approved
experiment). With a neutrino line as provided by $^7$Be
electron-capture in the sun, unique and unambiguous tests of neutrino
oscillation models can be carried out if one knows both the
charged-current and the neutral current reaction rates~\cite{lines}.

I believe we have  calculated the flux of $p$-$p$ neutrinos produced in the
sun to an accuracy of $\pm 1$\%. Unfortunately, we do not yet have a
direct measurement of this flux. The gallium experiments only tell us
the rate of capture of all neutrinos with energies above $0.23$ MeV.

The most urgent need for solar neutrino research is to develop a
practical experiment to measure directly the $p$-$p$ neutrino flux and
the energy spectrum of electrons produced by target interactions with
$p$-$p$ neutrinos.  Such an experiment can be used to test the precise
and fundamental standard solar model prediction of the $p$-$p$ neutrino flux.
Moreover, the currently favored neutrino oscillation solutions all
predict a strong influence of oscillations on the low-energy flux of
$\nu_e$.

\begin{figure}[!h]
\centerline{\psfig{figure=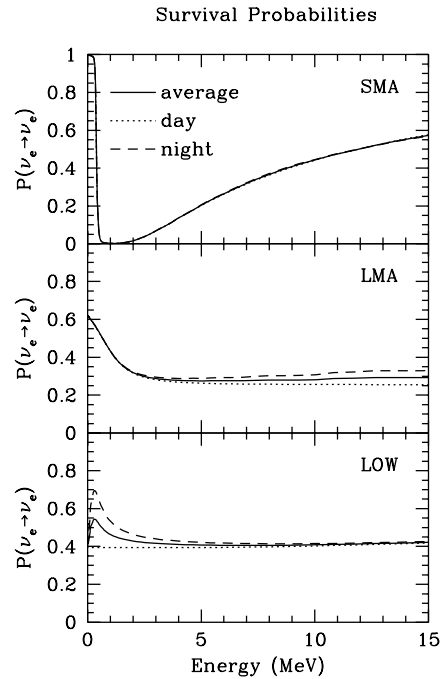,width=3.9in}}
\vglue-.6in
\caption[]{Survival probabilities for MSW solutions.
The figure presents the yearly-averaged 
survival probabilities for an electron neutrino that is  created
in the sun to remain an electron neutrino 
 upon arrival at the SuperKamiokande
detector.  There are only slight
differences between the computed regeneration probabilities for
the detectors located at the positions of Super-Kamiokande, 
SNO and the Gran Sasso
Underground Laboratory.
The full line refers to 
the average
survival probabilities computed  taking
into account regeneration in the earth and the dotted line refers to
calculations for the daytime 
that do not include regeneration. 
The dashed line includes regeneration at night.
This is Fig.~9 of the 1998 analysis by Bahcall, Krastev, and
Smirnov~\cite{bks98}. 
\label{fig:survival}}
\end{figure}

Figure~\ref{fig:survival} shows the calculated neutrino survival
probability as a function of energy for three global best-fit MSW
oscillation solutions.  You can see directly from this figure why we
have to have accurate measurements for the $p$-$p$ and $^7$Be neutrinos:
the currently favored solutions exhibit their most characteristic and
strongly energy dependent features below $1$ MeV.  In all of these
solutions, the survival probability shows a dramatic increase with
energy below $1$ MeV, whereas in the region above $5$ MeV (accessible
to SuperKamiokande and to SNO) the energy dependence of the survival
probability is at best modest.
 
The $p$-$p$ neutrinos are the gold ring of solar neutrino astronomy.
Their measurement will constitute a simultaneous and critical test of
stellar evolution theory and of neutrino oscillation solutions.

The most exciting result of this workshop for me has been the
possibility discussed here of a synergistic experiment involving a
huge (megaton?) nucleon decay detector in which an inner region is
reserved for solar neutrino experiments(see, for example, the talks by
Jung, Nakahata, and Ypsilantis at this workshop). The most
straightforward solar neutrino experiments that could be carried out
with this detector would be precision measurements of the temporal
dependences of the relatively high-energy $^8$B neutrinos. One could
measure with such a detector the zenith-angle dependence of the solar
neutrino-event rate (the generalization of the day-night difference)
and the seasonal dependence (generalization of the winter-summer
difference).  The design could relax somewhat the precise requirements
for energy calibration and for energy resolution used for the
SuperKamiokande and SNO experiments and concentrate instead on limiting the
systematic uncertainties in the detector that could contribute to
the error budget in the day-night or seasonal dependences.  After three
years of very careful measurements, the SuperKamiokande experiment
has, as we have heard at this conference, about a $2\sigma$ result for
the day-night difference. They do not yet have the statistics to
report a meaningful measurement of the full zenith-angle dependence or
the seasonal dependence. The predicted temporal effects are small,
generally of order a percent, with the currently favored neutrino
oscillation solutions.

Nature has provided us with many different baselines and with many
different matter column densities with which to do Very Long Baseline
(VLB) studies of neutrino oscillations. The earth-sun distance varies
continuously during the year between $1.496(1.0 \pm 0.017)10^{13}$ cm
and the column density through the earth to a terrestrial detector
varies from $0 {\rm~ gm~cm^{-2}}$ during the day to more than
$10^{9}{\rm ~gm~cm^{-2}}$ at night.

A solar neutrino detector ten or more times the volume of the current
SuperKamiokande experiment, as discussed in concept at this workshop,
could measure precisely the results of many different VLB neutrino
oscillation experiments.  This would be a fantastic `Smoking Gun'
detector.  I have a hard time sitting down when imagining such an
exciting possibility.

\section*{Acknowledgments}

I acknowledge support from NSF grant \#PHY95-13835.

\end{document}